\documentclass[prl,aps,twocolumn]{revtex4}

\usepackage{bm}
\usepackage{amsmath}
\usepackage{amssymb}

\def\unity{\mbox{\small 1} \!\! \mbox{1}}

\begin{document}

\date{\today}

\thispagestyle{empty}

\noindent
{\bf Comment on: ``Measuring a Photonic Qubit without Destroying It''}

\bigskip

\noindent
Recently, Pryde {\em et al}.\ reported the demonstration of a quantum
non-demolition (QND) scheme for single-photon polarization states with
linear optics and projective measurements \cite{pryde04}. In this
experiment, a single photon with a specific polarization in the signal
mode interacts on a beam splitter with a second polarized photon in
the meter mode. A destructive polarization measurement of the meter
photon then sometimes reveals the polarization of the signal photon
{\em without} the need for direct detection of the signal mode. This
allows the signal photon to propagate freely. Hence the interpretation
of this experiment as a single-photon QND measurement.  

To give a quantitative characterization of their QND scheme, Pryde
{\em et al}.\ introduced a {\em measurement fidelity} $F_M$, which
measures the overlap between the signal input and the measurement
distributions. This fidelity is based on the probabilities $P_{sm} =
P_{HH}$, $P_{HV}$, $P_{VH}$, and $P_{VV}$. Here, $P_{jk}$ is the
probability of finding a $j$-polarized photon in the signal mode, and
a $k$-polarized photon in the meter mode. In other words, $F_M$ is
determined solely by {\em coincidence counting}.   

However, for the protocol to work in true QND fashion, the signal
photon should propagate freely after the measurement. This means that
the proper fidelity measure of a QND protocol cannot be based on the
coincidence probabilities $P_{sm}$ alone: Using only coincidence
counting {\em necessarily} implies destructive photo-detection of 
the signal mode. This is incompatible with the definition of a
quantum non-demolition measurement. A proper measurement fidelity must
take into account $P_{k0}$ and $P_{0k}$, where 0 denotes the absence
of a detector count and $k\in\{ H,V\}$.  

Physically, when the circuit is operated in proper QND fashion, the signal
mode is not detected. This means that we have only the output of the
meter mode to tell us what the polarization state of the signal mode
is. But when the detectors have imperfections (low quantum efficiency
and lack of single-photon resolution), the meter-mode detection might
tell us there was only one horizontally polarized photon, when in fact
there was a second photon that failed to trigger the detector. In that
case, the signal mode is in the vacuum state, while we believe it has
a horizontally polarized photon. The probability that we mistake the
output vacuum for a horizontally polarized photon is given by $1 -
F_{QND}$, where we define the fidelity of the QND device $F_{QND}$ as
the overlap between the ideal output state and the physical output state
when photo-detection of the signal mode is omitted (for a detailed
discussion on the interpretation of the fidelity, see Ref.\
\cite{kok00}). This leads to $F_{QND} = {\rm Tr} \left[\hat{E}_k^{(1)}
  \hat{E}_{\perp k}^{(0)} \otimes |k\rangle_s \langle
  k|\;\hat{\rho}_{sm}\right]$, where $\hat{\rho}_{sm}$ is the density
operator of the state before detection, $|k\rangle$ is a single-photon
polarization state, and $\hat{E}_k^{(\ell)}$ is the Positive Operator
Valued Measure (POVM) that models the (imperfect) detection of $\ell$
photons with polarization $k$ in the meter mode.  

According to Pryde {\em et al}., the photo-detectors can distinguish
only between the vacuum state and non-vacuum states. Such detectors
cannot tell the difference between one and two photons. Furthermore,
the detectors have a probability $\zeta < 1$ of detecting a photon
\cite{pryde04}. The POVMs that correspond to such detectors are
derived in Ref.\ \cite{kok00}, and can be written as   
\begin{eqnarray}\nonumber
 \hat{E}_k^{(0)} &=& |0\rangle_k\langle 0| + (1-\zeta) |1\rangle_k
 \langle 1| + (1-\zeta)^2 |2\rangle_k\langle 2| \cr
 \hat{E}_k^{(1)} &=& \zeta |1\rangle_k\langle 1| + \zeta(2-\zeta)
 |2\rangle_k\langle 2| \; .
\end{eqnarray}
where $|n\rangle_k\langle n|$ is the projector onto the $n$-photon
Fock state in polarization mode $k$, and $\hat{E}_k^{(0)} +
\hat{E}_k^{(1)} = \unity$ on the truncated Fock space $\{ |0\rangle_k,
|1\rangle_k, |2\rangle_k\}$. The fidelity of the QND circuit is then
given by   
\begin{equation}
 F_{QND} = \frac{1}{2-\zeta}\; .
\end{equation}
The non-post-selected fidelity $F_{QND}$ with a typical detector
efficiency of $\zeta = 65$\% is approximately 0.74. This is
significantly less than the fidelity $F_M > 0.99$, measured by Pryde
{\em et al}. 

The fidelity $F_M$ is only meaningful when the circuit is conditioned
on coincidence counting, and this mode of operation is inconsistent
with a QND measurement. To characterize the QND mode of operation, a
different fidelity measure such as $F_{QND}$ must be used. With current
detectors, it is not clear whether $F_{QND}$ can be made sufficiently
large for quantum information processing \cite{knill04}.

\bigskip

\noindent
Pieter Kok and William J.\ Munro

Hewlett Packard Laboratories,

Filton Road, Stoke Gifford, Bristol BS34 8QZ, UK\\

\noindent 
{\small Date: \today }

\bibliography{po}

\end{document}